# Higher-order Bloch spheres: A generalized representation of electron spin states with azimuthal phase factor

Sota Sato[1], Toshiki Matsumoto[1], Yuichiro Nakano[1], Jun Ishihara[2], Katsuhiko Miyamoto[1, 3], Takashige Omatsu[1, 3], and Ken Morita[1, 3]

[1]*Graduate School of Engineering, Chiba University, Chiba, 263-8522, Japan*

[2]*Department of Applied Physics, Tokyo University of Science, Tokyo 125-8585, Japan*

[3]*Molecular Chirality Research Center, Chiba University, Chiba, 263-8522, Japan*

Abstract

Using the similarity between spin states on the Bloch sphere (BS) and polarization states on the Poincaré sphere (PS), we construct higher-order spin states on the higher-order BS corresponding to higher-order polarization states of photons on the higher-order PS. We investigate the time evolution of higher-order spin states in a magnetic field and establish an extended form of the conventional Larmor precession. The results provide insights on coherent transfer from extended photons to extended spin qubit systems with spin and orbital angular momenta and the operation of extended spin qubits.

**Corresponding author:** Ken Morita

Graduate School of Engineering, Chiba University,

Yayoi-cho 1-33, Inage-ku, Chiba 263-8522, Japan

Tel: +81-43-290-3360, Fax: +81-43-290-3360, E-mail: morita@chiba-u.jp

Photon and electron spins possessing a spin angular momentum (SAM) of ±1 are representative qubits in quantum information and communication technology [1,2]. The polarization states of photons (PSPs) are particularly suitable for reliable quantum information transport [3-5], whereas electron spins in semiconductor hosts are preferred for quantum information storage and processing [6-9]. A quantum information interface connecting different types of qubits is necessary to properly arrange these qubits. [10] The quantum media conversion between photon and electron spin enables the seamless deployment of quantum information transport, storage, and processing, which will help effectively exploit next-generation quantum communication technologies.

Similar to the PSP on the Poincaré sphere (PS), all spin states of electrons (SSEs) can be represented as superpositions of up- (north pole) and down-spins (south pole) on the Bloch sphere (BS). The PSP and SSE are equivalently represented on the PS and BS by employing the well-known Stokes and Bloch vectors. Both the PS and BS, which are spans of two basis vectors, belong to the equivalent special unitary group [Hilbert space SU(2)]. Furthermore, they are connected via selection rules in a semiconductor quantum well (QW) [10], which enables coherent transfer from a PSP to an SSE [11]. The utilization of light-hole (LH) excitation with a V-shaped three-level system in semiconductor QWs is key to realizing coherent transfer with SAM preservation [10,11].

Vector vortex beams, which possess a cylindrical symmetry of polarization arising from orbital angular momentum (OAM) [12], have attracted considerable attention in optical trapping [13] and terabit high-speed optical communications [14]. The polarization state of vector vortex beams can be represented on a higher-order PS, in which right- and left-handed circularly polarized OAM beams have the same topological charge but opposite helicity [15]. Recently, a more generalized higher-order PS was proposed by replacing the states at orthogonal poles with two arbitrary OAM beams with different topological charges [16]. Furthermore, the direct product of PSP with any orthogonal SAM and OAM allows for the generation of various vector vortex beams with different polarization states depending on the azimuthal angle [17,18]. Vector vortex beams, henceforth referred to as higher-order PSP, carry both SAM and OAM, even in a single-photon state [19,20]. Because the OAM can take any integer, the higher-order PSP is represented as a quantum state in high-dimensional Hilbert space [21].

Quantum state conversion from higher-order PSP with SAM and OAM to an electron system is an exciting topic regarding high-density quantum media conversion with a high degree of freedom and high-dimensional quantum operations for higher-order quantum entanglement. To date, optical excitation of electron systems in semiconductors by higher-order PSP has been demonstrated both theoretically [22,23] and experimentally [24-27]. However, these studies pertain to optical transitions that necessarily produce electron–hole entangled pairs without involving coherent transfer. To simultaneously clarify the coherent transfer of both SAM and OAM from the higher-order PSP to the electron system, it is necessary to construct an extended BS space, which is described by the direct product space of SAM and OAM.

In this letter, we represent extended states, referred to as a higher-order SSE on the higher-order BS, considering the similarity of a higher-order PSP on a higher-order PS. We consider coherent transfer using a V-shaped three-level system in the photon-to-spin as a starting point and construct an extended form of the spin system. A higher-order SSE possessing spin orientation with azimuth phase factor is determined from the SAM and OAM of photons. Furthermore, we investigate the time evolution of the higher-order SSE in an external magnetic field and establish an extended form of conventional Larmor precession.

We summarize the coherent transfer for quantum media conversion from PSPs on the PS [Fig. 1(a)] to SSEs on the BS [Fig. 1(b)] via LH excitation in a V-shaped three-level system [10,11] [Fig. 1(c)]. When an in-plane magnetic field is applied to a semiconductor QW, the energy degeneracy of LH is lifted and the eigenstates are reconfigured as $|\pm x\rangle_{LH} = (|\uparrow\rangle_{LH} \pm |\downarrow\rangle_{LH})/\sqrt{2}$, where $|\uparrow\rangle$ and $|\downarrow\rangle$ correspond to the up- and down-spin states, respectively. The energy degeneracy of the electrons is maintained with a small $g$-factor. The optical transition between $|-x\rangle_{LH}$ and degenerated electron states leads to the transfer of the PSP of $\alpha|R\rangle + \beta|L\rangle$ to $(\alpha|\uparrow\rangle_e + \beta|\downarrow\rangle_e) \otimes |-x\rangle_{LH}$, where $|R\rangle$ and $|L\rangle$ correspond to the right- and left-handed circularly polarized states, respectively. The direct product of the electron and LH states is physically equivalent to being unentangled, and such a transition allows coherent transfer from PSP to SSE.

The state vectors for PSP and SSE on the PS and BS connected by the above transition can be

expressed by the spherical coordinate $(\theta_\xi, \varphi_\xi)$ as

$$|\psi^\xi\rangle = \cos\frac{\theta_\xi}{2}|N^\xi\rangle + e^{i\varphi_\xi}\sin\frac{\theta_\xi}{2}|S^\xi\rangle, \qquad (1)$$

with respect to the orthogonal north and south poles $\{|N^\xi\rangle, |S^\xi\rangle\}$, where $\xi = P, B$ represent PS and BS, respectively. For PS, $\{|N^P\rangle, |S^P\rangle\}$ are replaced with right- and left-handed circularly polarized states $\{|R\rangle, |L\rangle\}$ [Fig. 1(a)], whereas for BS, $\{|N^B\rangle, |S^B\rangle\}$ are replaced with orthogonal up- and down-spin states $\{|\uparrow\rangle, |\downarrow\rangle\}$ [Fig. 1(b)]. Eq. (1) yields Stokes and Bloch vectors (see Supplemental Material S1 [28]).

Next, we describe the higher-order PSP and SSE. Henceforth, replace $\{\xi, \xi'\} = \{P, HP\}$ for the higher-order PSP, whereas $\{\xi, \xi'\} = \{B, HB\}$ is replaced for the higher-order BS. We first focus on higher-order PSP and subsequently present a higher-order SSE. First, we define the orthogonal basis $|\lambda^{\xi+}\rangle$ and $|\lambda^{\xi-}\rangle$ on PS as

$$|\lambda^{\xi+}\rangle = \cos\frac{\theta^\lambda_\xi}{2}|N^\xi\rangle + e^{i\varphi^\lambda_\xi}\sin\frac{\theta^\lambda_\xi}{2}|S^\xi\rangle, \quad |\lambda^{\xi-}\rangle = \sin\frac{\theta^\lambda_\xi}{2}|N^\xi\rangle - e^{i\varphi^\lambda_\xi}\cos\frac{\theta^\lambda_\xi}{2}|S^\xi\rangle, \qquad (2)$$

where $(\theta^\lambda_\xi, \varphi^\lambda_\xi)$ are the spherical coordinates [Fig. 2(a)]. The generalized higher-order PS has an orthogonal basis $|\chi^{\xi'+}_\ell\rangle = \exp(i\ell\phi)|\lambda^{\xi+}\rangle$ and $|\chi^{\xi'-}_m\rangle = \exp(im\phi)|\lambda^{\xi-}\rangle$ [Fig. 2(b)], where $\phi$ represents the azimuthal coordinate in real space and $(\ell, m)$ is the topological charge, which is proportional to the OAM of light. This allows the expression of the north pole $|N^{\xi'}_{\ell,m}\rangle$ and south pole $|S^{\xi'}_{\ell,m}\rangle$ on the generalized higher-order PS as

$$|N^{\xi'}_{\ell,m}\rangle = \cos\frac{\theta^\lambda_\xi}{2}|\chi^{\xi'+}_\ell\rangle + \sin\frac{\theta^\lambda_\xi}{2}|\chi^{\xi'-}_m\rangle,$$
$$|S^{\xi'}_{\ell,m}\rangle = e^{-i\varphi^\lambda_\xi}\left(\sin\frac{\theta^\lambda_\xi}{2}|\chi^{\xi'+}_\ell\rangle - \cos\frac{\theta^\lambda_\xi}{2}|\chi^{\xi'-}_m\rangle\right), \qquad (3)$$

with respect to the above-defined orthogonal basis $|\chi^{\xi'+}_\ell\rangle$ and $|\chi^{\xi'-}_m\rangle$. Finally, as in Eq. (1), the generalized higher-order PSP $|\psi^{\xi'}_{\ell,m}\rangle$ on the higher-order PS can be described by spherical coordinates $(\theta_{\xi'}, \varphi_{\xi'})$ as

$$|\psi_{\ell,m}^{\xi\prime}\rangle = \cos\frac{\theta_{\xi\prime}}{2}|N_{\ell,m}^{\xi\prime}\rangle + e^{i\varphi_{\xi\prime}}\sin\frac{\theta_{\xi\prime}}{2}|S_{\ell,m}^{\xi\prime}\rangle. \quad (4)$$

Note that after considering any new orthogonal bases $|\chi_\ell^{\xi\prime+}\rangle$ and $|\chi_m^{\xi\prime-}\rangle$, $|\psi_{\ell,m}^{\xi\prime}\rangle$ can be represented by its superposition as $|\psi_{\ell,m}^{\xi\prime}\rangle = \alpha|\chi_\ell^{\xi\prime+}\rangle + \beta|\chi_m^{\xi\prime-}\rangle$. Nevertheless, we used the orthogonal set of north $|N_{\ell,m}^{\xi\prime}\rangle$ and south $|S_{\ell,m}^{\xi\prime}\rangle$ poles for the general expression of $|\psi_{\ell,m}^{\xi\prime}\rangle$, as in Eq. (4), owing to the following requirements: First, for the state of higher-order PS ($\xi\prime = HP$) for $(\ell,m) = (0,0)$, $\phi = 0$ should be identical to the state of PS; hence, Eq. (4) must be extendable to the general form of Eq. (1). Second, the axes of the higher-order PS ($s_1^{HP}, s_2^{HP}, s_3^{HP}$) and higher-order BS ($s_x^{HB}, s_y^{HB}, s_z^{HB}$) are fixed with respect to the real space. Pauli spin matrices $\hat{\sigma}^\xi = (\hat{\sigma}_{0,I}^\xi, \hat{\sigma}_{1,x}^\xi, \hat{\sigma}_{2,y}^\xi, \hat{\sigma}_{3,z}^\xi)$ and their expected values are essential when considering the polarization and spin orientation of the higher-order PSP and SSE in real space. The azimuth-dependent polarization of the higher-order PSP can be described by a Stokes vector:

$$\langle s_1^P\rangle = \langle\psi_{\ell,m}^{HP}|\hat{\sigma}_1^P|\psi_{\ell,m}^{HP}\rangle, \quad \langle s_2^P\rangle = \langle\psi_{\ell,m}^{HP}|\hat{\sigma}_2^P|\psi_{\ell,m}^{HP}\rangle, \quad \langle s_3^P\rangle = \langle\psi_{\ell,m}^{HP}|\hat{\sigma}_3^P|\psi_{\ell,m}^{HP}\rangle \quad (5)$$

(see Supplemental Material S2 [28]). Fig. 2(c) shows the azimuth-dependent polarization of the higher-order PSP at representative points ($|H_{-1}^{HP}\rangle$, $|V_1^{HP}\rangle$, $|N_{-1,1}^{HP}\rangle$, $|S_{-1,1}^{HP}\rangle$, $|D_{-1,1}^{HP}\rangle$, $|A_{-1,1}^{HP}\rangle$) on the higher-order PS, assuming $(\theta_P^\lambda, \varphi_P^\lambda) = (\pi/2, 0)$ and $(\ell, m) = (-1, 1)$. Note that the orthogonal bases $|\chi_{-1}^{HP+}\rangle$ and $|\chi_1^{HP-}\rangle$ correspond to $|H_{-1}^{HP}\rangle$ and $|V_1^{HP}\rangle$, respectively. As shown, the polarization of $|\psi_{-1,1}^{HP}\rangle$ changes depending on $\phi$, except for the orthogonal bases $|\chi_{-1}^{HP+}\rangle(=|H_{-1}^{HP}\rangle)$ and $|\chi_1^{HP-}\rangle(=|V_1^{HP}\rangle)$. Thus, the higher-order PS is determined by the choice of orthogonal bases $|\chi_\ell^{HP+}\rangle$ and $|\chi_m^{HP-}\rangle$, which enables us to create various polarization states in which circular and linear polarization components can coexist depending on $\phi$.

Based on the correspondence between the PS and BS, as described in Eq. (1), we construct a higher-order BS by replacing $\xi$ with $B$ and $\xi'$ with $HB$ in Eqs. (2)-(4). Similar to the case of the higher-order PS [Fig. 2(b)], the points of the orthogonal basis ($|\chi_\ell^{HB+}\rangle$ and $|\chi_m^{HB-}\rangle$), north and south poles ($|N_{\ell,m}^{HB}\rangle$ and $|S_{\ell,m}^{HB}\rangle$), and generalized higher-order SSE $|\psi_{\ell,m}^{HB}\rangle$ on the higher-order BS are shown in Fig. 3(b). Similar to the case of higher-order PS, $|\chi_\ell^{HB+}\rangle$ and $|\chi_m^{HB-}\rangle$ are multiplied by $\exp(i\ell\phi)$ and $\exp(im\phi)$ for $|\lambda^{B+}\rangle$ and $|\lambda^{B-}\rangle$ on the BS [Fig. 3(a)]. To further reveal the higher-order SSE as observable quantities, the $\phi$-dependent spin orientation, that is, the expectation value of each spin component, is calculated using the Bloch vector [similar to Eq. (5)] expressed as

$$\langle s_x^B\rangle = \langle\psi_{\ell,m}^{HB}|\hat{\sigma}_x^B|\psi_{\ell,m}^{HB}\rangle,\ \langle s_y^B\rangle = \langle\psi_{\ell,m}^{HB}|\hat{\sigma}_y^B|\psi_{\ell,m}^{HB}\rangle,\ \langle s_z^B\rangle = \langle\psi_{\ell,m}^{HB}|\hat{\sigma}_z^B|\psi_{\ell,m}^{HB}\rangle \tag{6}$$

(see Supplemental Material S2 [28]). The spin orientation of the higher-order SSE corresponding to the higher-order PSP in Fig. 2(c) is calculated by substituting $(\theta_B^\lambda,\varphi_B^\lambda) = (\pi/2, 0)$ and $(\ell, m) = (-1, 1)$ into Eq. (6); the results are presented in Fig. 3(c). Similar to the polarization of $|\psi_{-1,1}^{HP}\rangle$ [Fig. 2(c)], the electron spin orientation of $|\psi_{-1,1}^{HB}\rangle$ changes depending on $\phi$, except for $|\chi_{-1}^{HB+}\rangle(=|H_{-1}^{HB}\rangle)$ and $|\chi_1^{HB-}\rangle(=|V_1^{HB}\rangle)$. In general, the individual electron spins point in the same orientation, which coincides with $|\lambda^{B+}\rangle$ and $|\lambda^{B-}\rangle$ for $|\chi_\ell^{HB+}\rangle$ and $|\chi_m^{HB-}\rangle$, respectively. For $|\psi_{l,m}^{HB}\rangle$ with $\phi = 0$ at $(\theta_{HB},\varphi_{HB})$ on the higher-order BS, the spin orientation coincides with the spin orientation at point $(\theta_B,\varphi_B) = (\theta_{HB},\varphi_{HB})$ on the BS. Furthermore, for $|\psi_{l,m}^{HB}\rangle$ with $\phi \neq 0$ at $(\theta_{HB},\varphi_{HB})$ on the higher-order BS, the spin orientation coincides with the circumferential state on the BS, which we refer to as the BS ring. The BS ring is a set of ring-shaped points where the BS intersects the perpendicular plane passing through point $(\theta_B, \varphi_B) = (\theta_{HB}, \varphi_{HB})$ in the direction of $|\lambda^{B+}\rangle$ and $|\lambda^{B-}\rangle$, as shown in Fig. 3(d). The spin orientation from $\phi = 0$ to $2\pi$ changes continuously on the BS ring, and its rotation period and direction are determined by $|\ell - m|$ and the sign of $\ell - m$, respectively. Fig. 3(e) shows the BS ring and spin states of $\phi = 0$ for the six states of the higher-order SSE in Fig. 3(c). The BS ring for $|N_{-1,1}^{HB}\rangle$, $|D_{-1,1}^{HB}\rangle$, $|S_{-1,1}^{HB}\rangle$, and $|A_{-1,1}^{HB}\rangle$ is shared on $s_x^B = 0$, and the BS ring for $|H_{-1}^{HB}\rangle$, $|V_1^{HB}\rangle$ converges to a single point, as $|H_{-1}^{HB}\rangle$, $|V_1^{HB}\rangle$ coincides with the orthogonal basis. The $\phi$-dependent spin orientation for $|N_{-1,1}^{HB}\rangle$, $|D_{-1,1}^{HB}\rangle$, $|S_{-1,1}^{HB}\rangle$, and $|A_{-1,1}^{HB}\rangle$ [Fig. 3(c)] corresponds to the spin orientation of the two rotations along the direction of the arrow on the BS ring [Fig. 3(e)] because $\ell - m = -2$. Thus, for higher-order SSE, the azimuth dependence of the spin orientation can be revealed by introducing a BS ring. The same idea can be applied to higher-order PSP. Defining the higher-order state vectors in photon and spin as in Eqs. (3) and (4) also yields higher-order Stokes [15] and Bloch vectors (see Supplementary Material S3 [28]). The equivalence of photon and spin systems is maintained in the higher- and lower-order cases. The discussion thus far shows that for the higher-order PS, the higher-order BS is determined by the choice of orthogonal basis, which enables us to create various spin orientations depending on $\phi$. We have developed a fundamental theory that assists in the generation of new quantum spin states with SAM and OAM in semiconductors.

An electron spin rotates around an external magnetic field on the BS, which is known as the Larmor precession. Hence, the spin is generally controlled by an external magnetic field. The control of spin by an external magnetic field corresponds to the polarization control of light by a wave plate in the low-order case. Next, we will describe the motion of a higher-order SSE on the higher-order BS under a static external magnetic field. As in the conventional SSE, the Larmor precession of the higher-order SSE can be developed using the time-dependent Schrödinger equation expressed as follows:

$$i\hbar \frac{\partial}{\partial t}|\psi_{\ell,m}^{HB}(t)\rangle = \hat{\mathcal{H}}|\psi_{\ell,m}^{HB}(t)\rangle, \tag{7}$$

where $\hat{\mathcal{H}}$ is the Hamiltonian for the Zeeman interaction, which is expressed as

$$\hat{\mathcal{H}} = \frac{1}{2}g\mu_B B n \cdot \hat{\sigma}, \tag{8}$$

where $g$ is the electron $g$-factor, $\mu_B$ is the Bohr magneton, and $B$ is the magnitude of the magnetic field directed at unit vector $n$. By solving Eq. (7), we obtain the solution for the well-known time evolution described by

$$|\psi_{\ell,m}^{HB}(t)\rangle = \exp\left(-i\frac{\omega t}{2}n \cdot \hat{\sigma}\right)|\psi_{\ell,m}^{HB}(0)\rangle \equiv \hat{R}_n(\omega t)|\psi_{\ell,m}^{HB}(0)\rangle, \tag{9}$$

where $\omega = g\mu_B B/\hbar$ is the precession frequency and $\hat{R}_n(\omega t)$ is the rotation operator representing the Larmor precession. By choosing the initial states as $|\psi_{\ell,m}^{HB}(0)\rangle = \alpha|\chi_\ell^{HB+}\rangle + \beta|\chi_m^{HB-}\rangle$, the time evolution of the higher-order SSE $|\psi_{\ell,m}^{HB}(t)\rangle$ on the higher-order BS can be described as

$$|\psi_{\ell,m}^{HB}(t)\rangle = \alpha|\chi_\ell^{HB+}(t)\rangle + \beta|\chi_m^{HB-}(t)\rangle, \tag{10}$$

where $|\chi_\ell^{HB+}(t)\rangle = \hat{R}_n(\omega t)|\chi_\ell^{HB+}\rangle$ and $|\chi_m^{HB-}(t)\rangle = \hat{R}_n(\omega t)|\chi_m^{HB-}\rangle$. Thus, the time evolution of $|\psi_{\ell,m}^{HB}(t)\rangle$ can be represented as a superposition of the orthogonal basis ($|\chi_\ell^{HB+}(t)\rangle$ and $|\chi_m^{HB-}(t)\rangle$)

that precess the magnetic field, as shown in Fig. 4(a). Here, the values of $\alpha$ and $\beta$ are time-independent, indicating that the relative positions of $|\psi_{\ell,m}^{HB}(t)\rangle$, $|\chi_\ell^{HB+}(t)\rangle$, and $|\chi_m^{HB-}(t)\rangle$ on the higher-order BS do not change. Therefore, $|\psi_{\ell,m}^{HB}(t)\rangle$ can be regarded as the extended SSE precessing around the external magnetic field on a time-varying higher-order BS represented by time-dependent orthogonal bases $|\chi_\ell^{HB+}(t)\rangle$ and $|\chi_m^{HB-}(t)\rangle$. The above discussion can be summarized as follows. As already shown, the higher-order BS depends on the choice of the orthogonal basis ($|\chi_l^{HB+}\rangle$ and $|\chi_m^{HB-}\rangle$). When an external magnetic field is applied, the basis of the higher-order BS is generally time-dependent ($|\chi_l^{HB+}(t)\rangle$ and $|\chi_m^{HB-}(t)\rangle$) because the direction of the external magnetic field and chosen basis differ in most cases. Thus, the higher-order BS is time-varying during the precession of state vector $|\psi_{l,m}^{HB}(t)\rangle$. In contrast, applying this concept to spin precession on the conventional BS, all states on the BS can be represented by a superposition of an arbitrary orthogonal basis on its BS. Hence, the BS is time-stationary during spin precession and independent of the direction of the external magnetic field and chosen basis. Consequently, the physical description of the time-evolution of the spin state vector on the higher-order and conventional BSs differs in terms of time-varying and time-stationary spherical rotations. Fig. 4(b) shows an example of the time evolution of the higher-order SSE on the time-varying higher-order BS when $|\psi_{\ell,m}^{HB}(t=0)\rangle$ starts at $(\theta_{HB}, \varphi_{HB}) = (0,0)$ and $(\theta_B^\lambda, \varphi_B^\lambda) = (\pi/2, 0)$, $(\ell, m) = (-1,1)$; that is, $|N_{-1,1}^{HB}\rangle$ in Fig. 3(c), and the magnetic field is applied in the $y$-direction in real space. The state vector of $|\psi_{\ell,m}^{HB}(t)\rangle$ rotates with the bases $|\chi_\ell^{HB+}(t)\rangle$ and

$|\chi_m^{HB-}(t)\rangle$ around the magnetic field, as shown on the left side of Fig. 4(b). $\phi$-dependent spin orientation of the corresponding states is shown on the right-hand side of Fig. 4(b). The motion of each spin follows the motion of the BS ring as it rotates around the external magnetic field. From the equivalence of photon and spin discussed in the previous section, the concept of the time-varying higher-order BS should enable understanding of the control of higher-order PSP by the waveplate angle and retardation.

In conclusion, the higher-order SSE on the higher-order BS corresponding to a higher-order PSP on the higher-order PS is presented, considering the similarity between SSE on the BS and PSP on the PS. The azimuth-dependent spin orientation of the higher-order SSE determined from the SAM and OAM of photons is calculated by the expectation value of each spin component given by the Pauli spin matrices. As in the polarization state in the higher-order PS, the spin state in the higher-order BS is determined by the choice of the orthogonal basis, and its orientation depends on the azimuth angle. The BS-ring is presented to easily understand the individual orientations of the spins in the higher-order SSE. The Larmor precession of the higher-order SSE in an external magnetic field was found to rotate on the time-varying higher-order BS, whose orthogonal basis depends on time. The spin precession features differ from conventional spin precession on a time-stationary BS. The equivalence of photon and spin systems is maintained in the higher- and lower-order cases, which allows for the realization of SAM and OAM coherent transfers in higher-order systems. These results provide clear guidelines, not only for coherent transfer from extended photons to extended spin qubit systems with SAM and OAM but also for the operation of extended spin qubits, which will be key aspects in future quantum technology.

**Acknowledgments**

We thank E. Abe, Y. Ishitani, and S. Iba for the constructive discussions. This work was partially supported by the Japan Society for the Promotion of Science (JSPS) (Grant No. 22H01981) and the Murata Science Foundation.

Figure captions

FIG. 1. (a) Stokes state vector $|\psi^P\rangle$ represented on the PS (magenta). (b) Bloch state vector $|\psi^B\rangle$ represented on the BS (cyan). Each state on the PS and BS is connected by coherent transfer via an LH excitation in a (c) V-shaped three-level system. The V-shaped three-level system consists of degenerate electron states and one eigenstate of non-degenerate LH states.

FIG. 2. (a) State vectors of the orthogonal bases $|\lambda^{P+}\rangle$ (purple) and $|\lambda^{P-}\rangle$ (green) on the PS. (b) Higher-order state vectors of $|\psi_{\ell,m}^{HP}\rangle$ (magenta), orthogonal bases $|\chi_\ell^{HP+}\rangle$ (purple) and $|\chi_m^{HP-}\rangle$ (green), and north $|N_{\ell,m}^{HP}\rangle$ (red point) and south $|S_{\ell,m}^{HP}\rangle$ (blue point) poles. $|\chi_\ell^{HP+}\rangle$, $|\chi_m^{HP-}\rangle$, and the north $|N_{\ell,m}^{HP}\rangle$ and south $|S_{\ell,m}^{HP}\rangle$ poles are the primary vectors that constitute the higher-order PS. (c) A higher-order PSP at representative points ($|H_{-1}^{HP}\rangle$, $|V_1^{HP}\rangle$, $|N_{-1,1}^{HP}\rangle$, $|S_{-1,1}^{HP}\rangle$, $|D_{-1,1}^{HP}\rangle$, and $|A_{-1,1}^{HP}\rangle$) on the higher-order PS assuming $(\theta_P^\lambda, \varphi_P^\lambda) = (\pi/2, 0)$ and $(\ell, m) = (-1, 1)$.

FIG. 3. (a) State vectors of the orthogonal bases $|\lambda^{B+}\rangle$ (purple) and $|\lambda^{B-}\rangle$ (green) on the BS. (b) Higher-order state vectors of $|\psi_{\ell,m}^{HB}\rangle$ (cyan), orthogonal basis $|\chi_\ell^{HB+}\rangle$ (purple) and $|\chi_m^{HB-}\rangle$ (green), and north $|N_{\ell,m}^{HB}\rangle$ (red point) and south $|S_{\ell,m}^{HB}\rangle$ (blue point) poles. $|\chi_\ell^{HB+}\rangle$, $|\chi_m^{HB-}\rangle$, and the north $|N_{\ell,m}^{HB}\rangle$ and south $|S_{\ell,m}^{HB}\rangle$ poles are the primary vectors that constitute the higher-order BS. (c) A higher-order SSE at representative points ($|H_{-1}^{HB}\rangle$, $|V_1^{HB}\rangle$, $|N_{-1,1}^{HP}\rangle$, $|S_{-1,1}^{HP}\rangle$, $|D_{-1,1}^{HP}\rangle$, $|A_{-1,1}^{HP}\rangle$) on the higher-order BS corresponding to Fig. 2(c). $\phi$-dependent spin orientation is obtained except for orthogonal bases $|H_{-1}^{HB}\rangle$ and $|V_1^{HB}\rangle$. Spin orientation at $\phi = 0$ (dark blue) in the higher-order SSE on the higher-order BS coincides with the spin orientation in the same coordinates on the BS. (d) The BS-ring (red) on the BS representing spin orientation from $\phi = 0$ to $2\pi$ in the higher-order SSE. (e) The BS-ring and spin states of $\phi = 0$ correspond to higher-order SSE in Fig. 3(c). The BS-ring (red) is on the $s_x^B = 0$ planes for $|N_{-1,1}^{HB}\rangle$, $|D_{-1,1}^{HB}\rangle$, $|S_{-1,1}^{HB}\rangle$, and $|A_{-1,1}^{HB}\rangle$, whereas it converges to a single red point for $|H_{-1}^{HB}\rangle$ and $|V_1^{HB}\rangle$.

FIG. 4. (a) Time-evolution of $|\psi_{\ell,m}^{HB}(t)\rangle$ (cyan) and the orthogonal bases $|\chi_\ell^{HB+}(t)\rangle$ (purple) and $|\chi_m^{HB-}(t)\rangle$ (green) which precess simultaneously around the external magnetic field (red) on the time-varying higher-order BS. (b) Time-evolution of $|\psi_{\ell,m}^{HB}(t)\rangle$ on the time-varying higher-order BS (left) and the corresponding higher-order SSE (right), where $|\psi_{\ell,m}^{HB}(0)\rangle$ starts from $(\theta_{HB}, \varphi_{HB}) = (0,0)$ under $(\theta_B^\lambda, \varphi_B^\lambda) = (\pi/2, 0)$, $(\ell, m) = (-1, 1)$, and the magnetic field is applied in the $y$-direction in

real space. Four states that circle on the higher-order BS and return to $|\psi_{\ell,m}^{HB}(0)\rangle$ are represented.

FIG. 1.

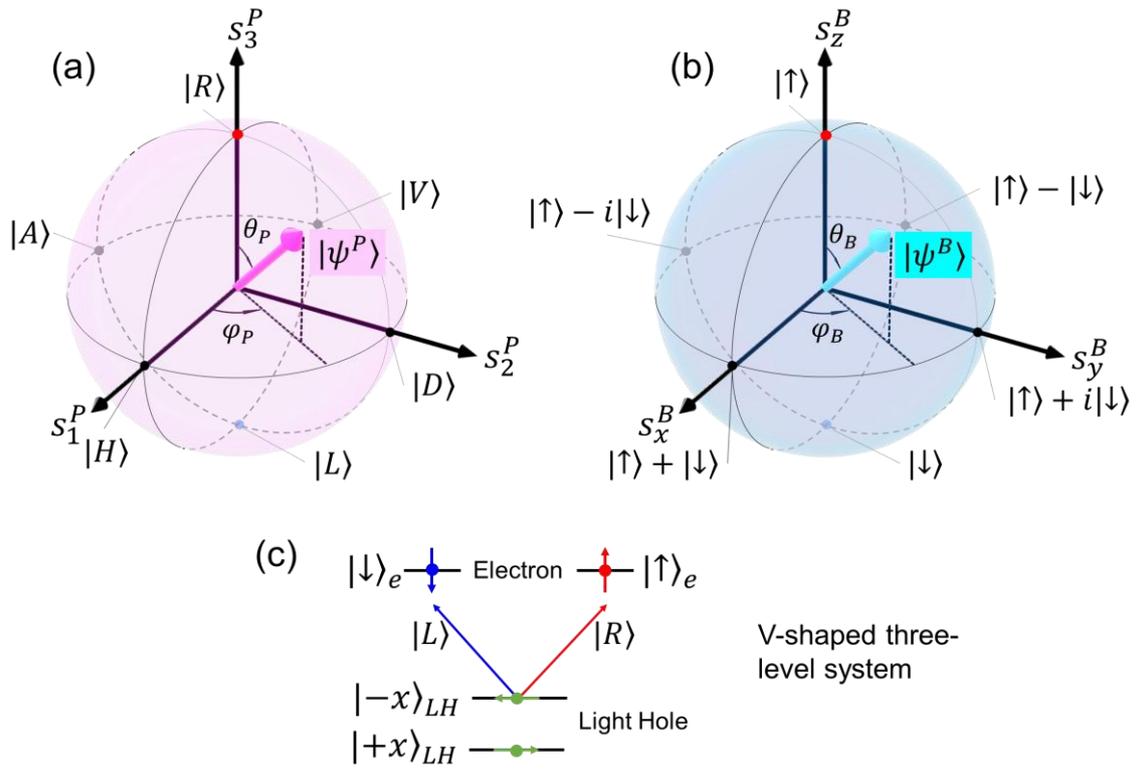

FIG. 2.

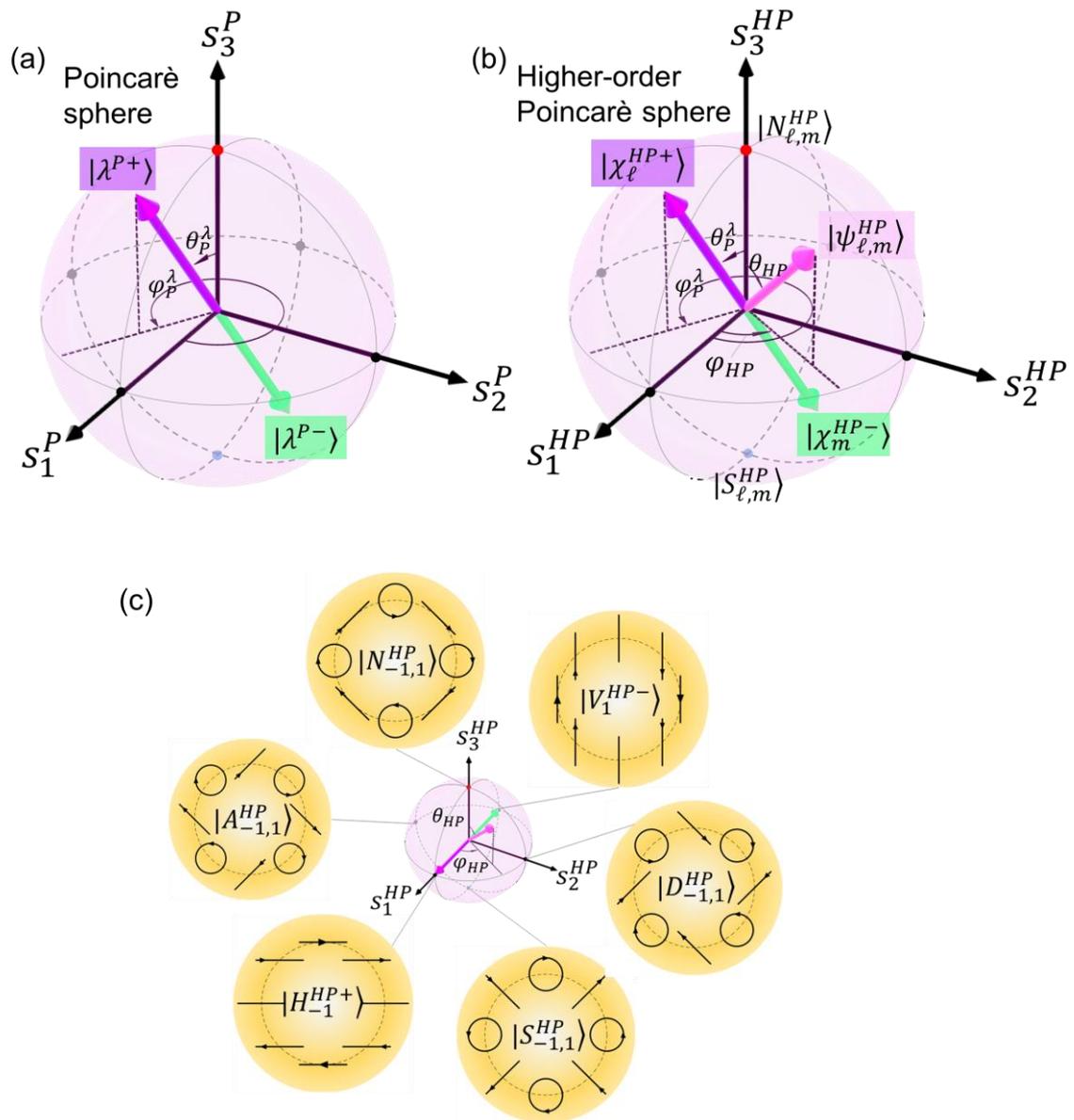

FIG. 3.

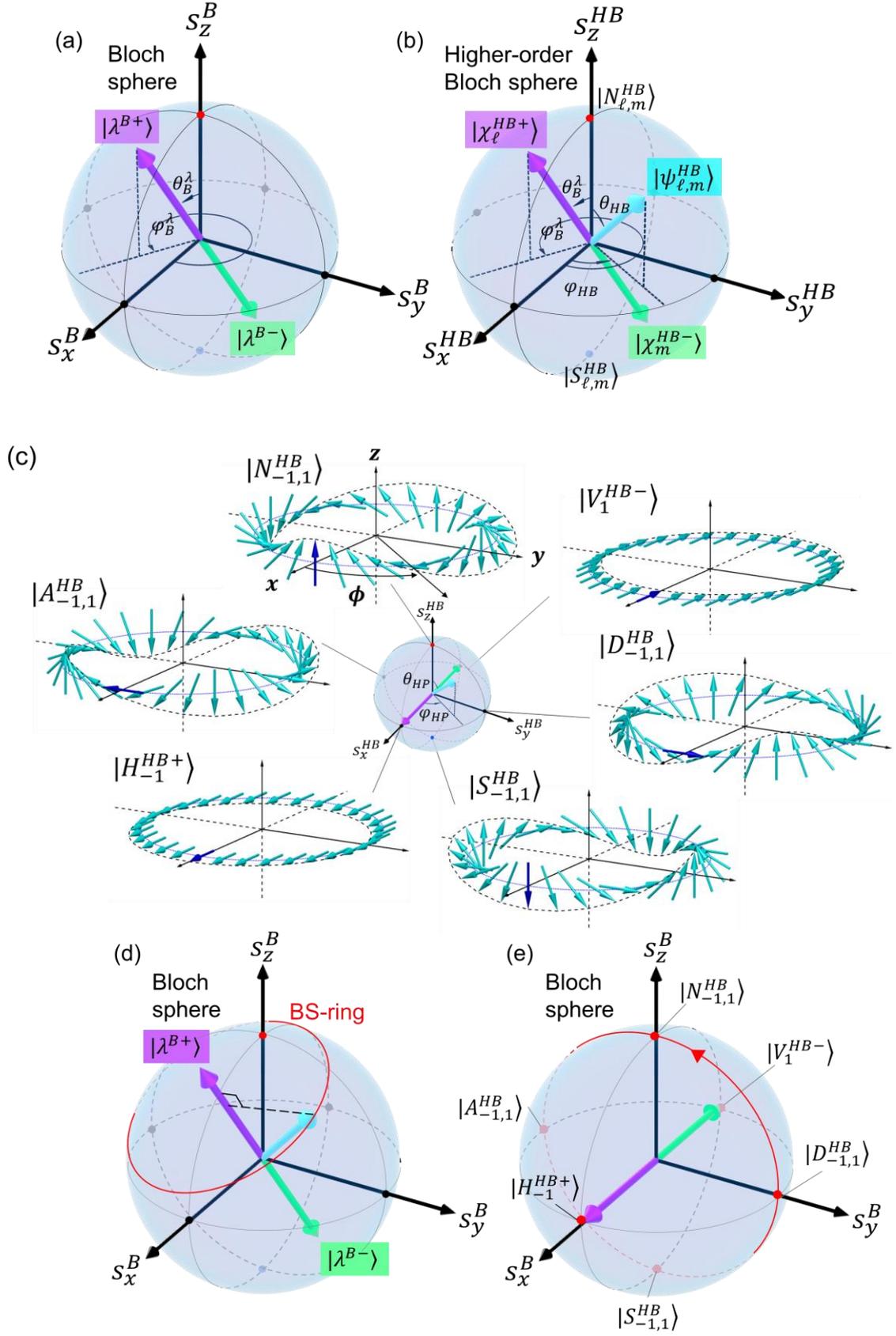

FIG. 4.

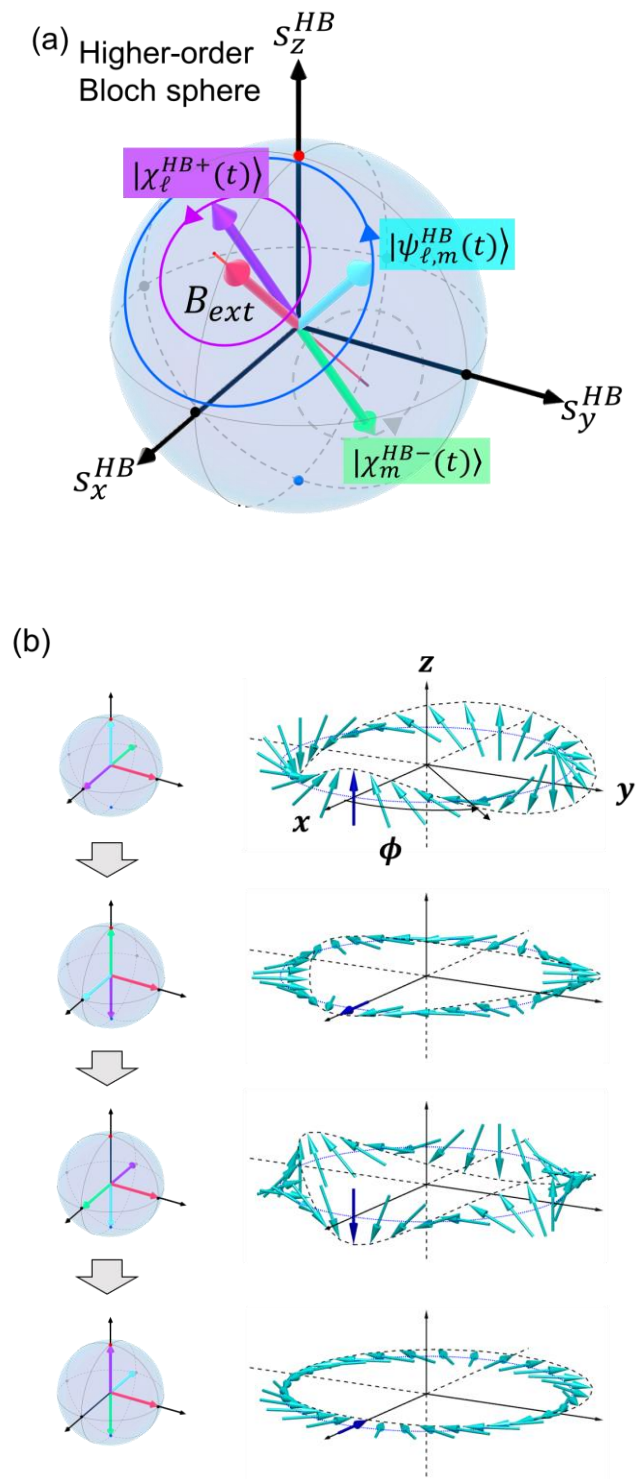